\documentclass[]{aastex701}
\usepackage{longtable}
\usepackage{subfig}
\usepackage{graphicx}
\usepackage{threeparttablex}
\usepackage{multirow}
\usepackage{lineno}

\begin{document}

\title{The Double-Burst Nature and Early Afterglow Evolution of Long GRB 110801A}

\author[0009-0000-9352-6447]{Qiu-Li Wang}
\affiliation{Purple Mountain Observatory, Chinese Academy of Sciences, Nanjing 210023, China}
\affiliation{School of Astronomy and Space Science, University of Science and Technology of China, Hefei 230026, China}
\email{qlwang@pmo.ac.cn}

\author[0000-0003-2915-7434]{Hao Zhou}
\affiliation{Purple Mountain Observatory, Chinese Academy of Sciences, Nanjing 210023, China}
\email{haozhou@pmo.ac.cn}

\author[0000-0002-8385-7848]{Yun Wang}
\affiliation{Purple Mountain Observatory, Chinese Academy of Sciences, Nanjing 210023, China}
\email{wangyun@pmo.ac.cn}

\author[0000-0002-9037-8642]{Jia Ren}
\affiliation{Purple Mountain Observatory, Chinese Academy of Sciences, Nanjing 210023, China}
\email{renjia@pmo.ac.cn}

\author[0000-0003-4977-9724]{Zhi-Ping Jin}
\affiliation{Purple Mountain Observatory, Chinese Academy of Sciences, Nanjing 210023, China}
\affiliation{School of Astronomy and Space Science, University of Science and Technology of China, Hefei 230026, China}
\email{jin@pmo.ac.cn}

\author[0000-0002-9758-5476]{Da-Ming Wei}
\affiliation{Purple Mountain Observatory, Chinese Academy of Sciences, Nanjing 210023, China}
\affiliation{School of Astronomy and Space Science, University of Science and Technology of China, Hefei 230026, China}
\email{dmwei@pmo.ac.cn}

\correspondingauthor{Zhi-Ping Jin, Da-Ming Wei}
\email{jin@pmo.ac.cn, dmwei@pmo.ac.cn}

\begin{abstract}

We present a comprehensive temporal and spectral analysis of the long-duration gamma-ray burst GRB 110801A, utilizing multi-band data from the Neil Gehrels Swift Observatory and ground-based telescopes. The $\gamma$-ray emission exhibits a distinct two-episode (``double-burst'') structure. Rapid follow-up observations in the optical and X-ray bands provide full coverage of the second burst. The optical light curve begins to rise approximately 135 s after the trigger, significantly preceding the second emission episode observed in X-rays and $\gamma$-rays at $\sim 320$ s. This chromatic behavior suggests different physical origins for the optical and high-energy emissions. Joint broadband spectral fitting (optical to $\gamma$-rays) during the second episode reveals that a two-component model, consisting of a power-law plus a Band function, provides a superior fit compared to single-component models. We interpret the power-law component as the afterglow of the first burst (dominating the optical band), while the Band component is attributed to the prompt emission of the second burst (dominating the high-energy bands). A physical synchrotron model is also found to be a viable candidate to explain the high-energy emission. Regarding the afterglow, the early optical light curve displays a sharp transition from a rise of $\sim t^{2.5}$ to $\sim t^{6.5}$, which is well-explained by a scenario involving both reverse shock (RS) and forward shock (FS) components. We constrain the key physical parameters of the burst, deriving an initial Lorentz factor $\Gamma_0 \sim 60$, a jet half-opening angle $\theta_j \sim 0.09$, and an isotropic kinetic energy $E_{\rm k,iso} \sim 10^{54.8}$ erg.

\end{abstract}

\keywords{Gamma-ray bursts (629)}


\section{Introduction} 
\label{sec:intro}
Gamma-ray bursts (GRBs) represent the most energetic explosions in the universe that originated from either the collapse of massive stars \citep{RevModPhys.76.1143, annurev:/content/journals/10.1146/annurev.astro.43.072103.150558} or the merger of compact binaries \citep{Abbott_2017}. Following a catastrophic event, an outflow is powered by the central engine and is likely collimated into a relativistic jet during propagation \citep{2022Galax..10...93S}. The nature of prompt emission can be explained by the energy dissipation in the internal shock \citep{1994ApJ...430L..93R, Kobayashi_1997,2014A&A...562A.100E} or magnetic dissipation \citep{2016MNRAS.459.3635B,2016ApJ...816L..20G,2017Natur.547..425T}, although its complex radiation mechanisms remain poorly understood. The most natural candidate is synchrotron radiation generated from energetic electrons. However, the theoretical synchrotron spectrum predicts photon indices $\alpha \sim-1.5$ in lower energy band expected for the fast cooling synchrotron model are softer than typical photon indices ($\bar{\alpha}\sim -1$), which is the so-called fast cooling problem\citep{10.1046/j.1365-8711.2000.03354.x}. To reconcile this inconsistency, several modifications to the canonical synchrotron framework have been suggested, such as additional inverse Compton scattering \citep{2009ApJ...703..675N, 2011A&A...526A.110D} or a decaying magnetic field \citep{2006ApJ...653..454P, 2014NatPh..10..351U,2014ApJ...780...12Z}. Another alternative interpretation is the presence of a sub-dominant quasi-thermal component (in addition to a non-thermal component)\citep{2011ApJ...727L..33G, 2014ApJ...795..155P,2021MNRAS.501.4974V,2025ApJ...987..129W}. The afterglow is powered by synchrotron radiation from the external forward shock or reverse shock, produced as the jet interacts with the circumburst medium \citep{1997ApJ...476..232M,1999MNRAS.306L..39M,1999ApJ...517L.109S,2005Natur.435..181B,2012ApJ...748...59G,2024NatAs...8..134A,2025ApJ...990..110W}.

The prompt emission of GRB 110801A in $\gamma$-ray shows two comparable energetic episodes separated by a quiescence time of roughly 200\,s like GRB 240529A \citep{2024ApJ...976L..20S}. The optical observation starts during the $\gamma$-ray's quiescent time and shows a sharp rise. The two isolated $\gamma$-ray episodes and the sharp transition in the optical rise imply the presence of "refreshed shock". The "refreshed shock" scenario is commonly invoked to explain the re-brightening behavior observed in late-time afterglows \citep{2014MNRAS.445.1625N,2020ApJ...899..105L,2023MNRAS.525.5224M, 2024ApJ...970..139S}. In this paper, we report the analysis of long GRB 110801A data obtained by the Neil Gehrels \emph{Swift} Observatory \citep{Gehrels_2004}. Adopting the procedure in \cite{2023NatAs...7.1108J} and \cite{2023ApJS..268...65Z}, we rebinned the ultraviolet/optical telescope \citep[UVOT, ][]{roming_swift_2005} data taken in the event mode into finer time bins. This allowed us to investigate the optical temporal behavior during the early stage of the burst. We conclude the optical flux is dominated by the earlier shock's afterglow emission, while the X-ray and $\gamma$-ray flux is dominated by the later shell's prompt emission. 

Here we present the observations and the reduction of the data described in Sec \ref{sec:observations}.The analysis of prompt emission and broadband spectral fitting results are detailed in Section \ref{sec:prompt}, while the afterglow analysis is addressed in Section \ref{sec:afterglow}. We conclude with a summary and discussion in Section \ref{sec:summary}.
Throughout this work, we adopt a standard flat cosmology parameterized by $H_{0}=67.4$ km s$^{-1}$Mpc$^{-1}$, $\Omega_{M}=0.315$ and $\Omega_{\Lambda}=0.685$ \citep{2020A&A...641A...6P}. Unless specified otherwise, all reported uncertainties reflect the 1$\sigma$ confidence interval.

%% The "ht!" tells LaTeX to put the figure "here" first, at the "top" next
%% and to override the normal way of calculating a float position.
%% The asterisk after "figure" tells the compiler to span multiple columns
%% if a two column style is selected.

\section{OBSERVATIONS AND DATA REDUCTIONS} \label{sec:observations}
\subsection{Observations} 
\label{subsec:subsection21}

At 19:49:42 UT on August 01, 2011  ($T_{0}$), GRB 110801A triggered The {\it Swift} burst alert telescope \citep[BAT, ][]{barthelmy_burst_2005,2011GCN.12228....1D}. Following the BAT localization,  {\it Swift} immediately slewed to the source and initiated follow-up observations. Then the X-ray telescope \citep[XRT, ][]{burrows_swift_xrt} began to observe the field at 19:51:21.9 UT, 99 seconds after
the BAT trigger. XRT found a bright, uncatalogued X-ray source located
at RA = 05h 57m 43.36s, DEC = +80$\,^\circ$ 57$\,\arcmin$ 17.2$\,\arcsec$ (J2000) with a 90\% uncertainty of 4.9$\,\arcsec$. UVOT began to observe the field of GRB 110801A about 108 seconds after the BAT trigger in the event mode. In addition, the afterglow counterpart of GRB 110801A was observed with the 10.4m GTC telescope revealing a continuum spectrum and narrow absorption lines, with a likely redshift of 1.858 \citep{2011GCN.12234....1C}. The burst was also detected by Konus-Wind \citep{2011GCN.12276....1S}. Since the Konus-Wind observed this GRB in the waiting mode, they only have 3 channel spectral data for the Konus-Wind which cover the energy range from 20 keV to 1.2 MeV. The wide energy ranges of BAT(15 keV to 150 keV) and Konus-Wind show that the time-averaged spectrum from $T_0-26$\,s to 
$T_0+65$\,s and from $T_0+332$\,s to $T_0+389$\,s include two time-separated episodes are best fit with a power-law with exponential cutoff model:
\begin{equation}\label{equ:R}
\frac{\mathrm{d} N}{\mathrm{d} E}  \sim E^{\alpha } \exp\left[-(2+\alpha)\frac{E}{E_{\rm peak}}\right]
\end{equation}
The best fit spectral parameters are: 
$\alpha= -1.7^{+0.1}_{-0.2}$ and $E_{\rm peak}=140^{+900}_{-60}$ keV. The energy fluence in the 15-1200 keV band calculated by a power-law with exponential cutoff model is $7.3^{+ 1.7}_{-0.9}\times 10^{-6} $\,$\rm erg\,/cm^{2}$.The best fit spectral parameters for the Band model fixing $\beta= -2.5$ are: $\alpha= -1.70_{-0.15}^{+0.22}$ , and $E_{peak}=140_{-50}^{+1270}$ keV. The isotropic energy $E_{iso}$ from 1 keV to 10 MeV band is $1.0^{+0.3}_{-0.2}\times10^{53}$ erg, using the Band spectrum model.

Furthermore, the optical three color (g, Rc and Ic) CCD camera attached to the Murikabushi 1m telescope of Ishigakijima Astronomical Observatory observed the field of GRB 110801A on 19:53:51 UT, about 4 min after the BAT trigger and obtained a series of images with exposure time of 60 seconds \citep{2011GCN.12233....1K}. 

\subsection{Data reduction} \label{subsec:subsection2}
We employed the HEASoft software package (version 6.35) to reduce the data obtained from{\it Swift} BAT, XRT and UVOT. The BAT light curve (15 - 350\,keV) from $T_0-100$\,s to $T_0+900$\,s was extracted with a time bin size of 1\,s and generated using standard procedures described in the {\it Swift}/BAT software guide\footnote{\url{https://www.swift.ac.uk/analysis/bat/index.php}}. The XRT source and background spectral files in both PC and WT modes the redistribution matrix, and ancillary response files were reduced with the online {\it Swift}/XRT data products generator\footnote{\url{https://www.swift.ac.uk/burst_analyser/}}\citep{ev2007, ev2009}. UVOT observed GRB 110801A in the V, B, U, W1 and WHITE bands for several epochs. We performed photometry on the level 2 UVOT image mode products using a standard aperture. For the event-mode observations, we extracted the WHITE-band and U-band light curves with fixed time bins of 50,s and 20,s, respectively, applying the same standard aperture. The UVOT photometries are listed in Table \ref{table:2}

\section{PROMPT EMISSION ANALYSIS} \label{sec:prompt}
\subsection{Broadband lightcurves of the prompt emission} \label{subsec:subsection31}
\begin{figure}[!hb] 
  \centering  
  \includegraphics[width=1\linewidth]{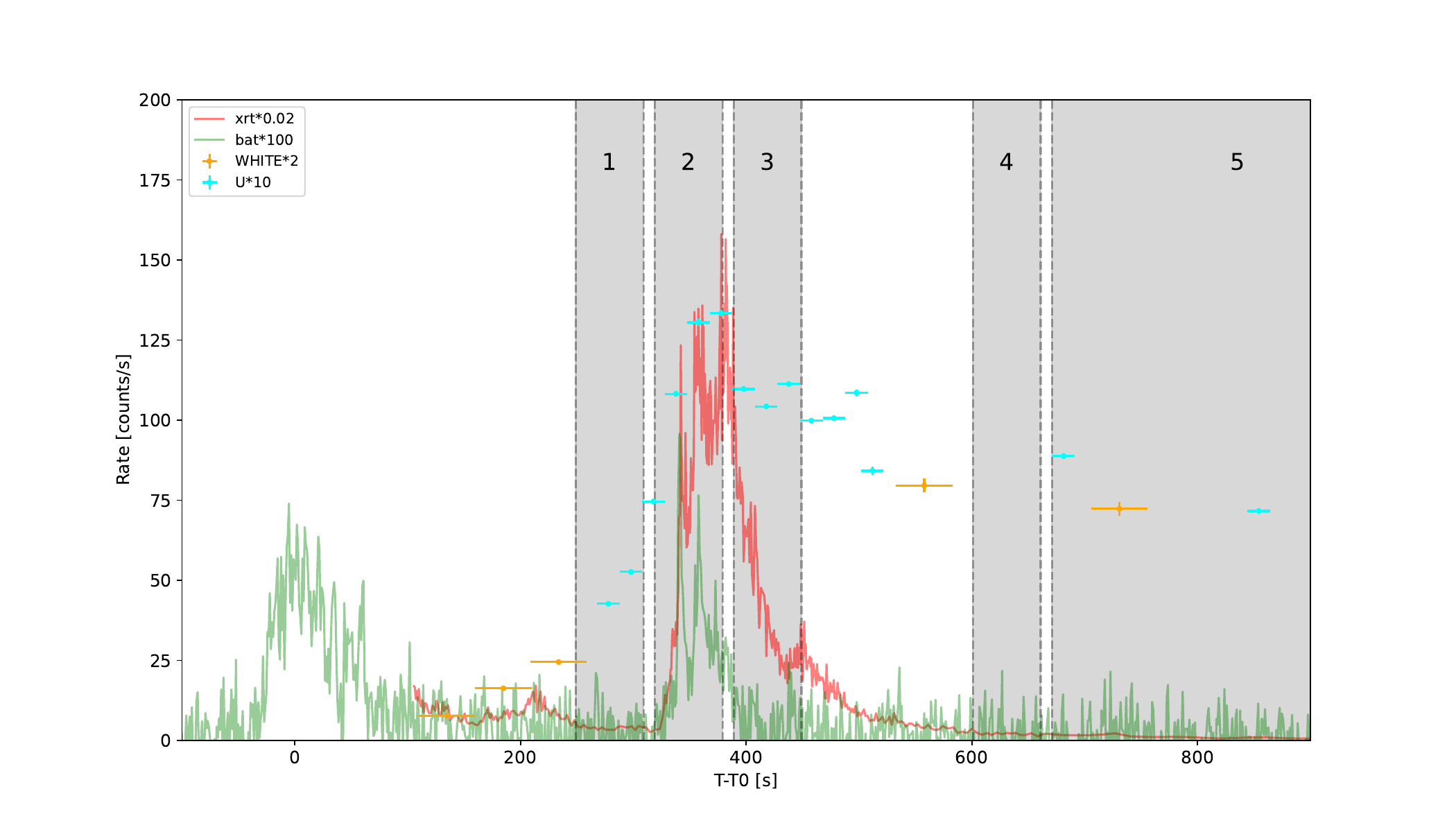}
  \caption{Multi-band emission light curves of GRB 110801A. Data are collected by {\it Swift} BAT (green), XRT (red), and UVOT. Note that the WHITE-band data are in orange and U-band data are in cyan. The vertical dashed lines represent the start and end times of the  intervals selected for the spectral analysis.}
  \label{fig:1}
\end{figure}

As shown in Figure \ref{fig:1}, the {\it Swift} BAT (15–350 keV) light curve exhibits two groups of several peaks from $T_0-50$\,s to $T_0+80$\,s and $T_0+320$\,s to $T_0+400$\,s with a quiescent period (SNR $<$ 3) at $\sim100-300$\,s. Initially, we interpret the first burst as a precursor emission, as it is common to observe a weaker precursor followed by a quiescent period preceding the main burst \citep{1995ApJ...452..145K, 2008ApJ...685L..19B, 2014ApJ...789..145H}. However, the peak flux of the first burst is comparable to that of the second, and the duration of the first burst is even longer than that of the second. These facts compel us to abandon this interpretation. These episodes have similar activity amplitudes, which would correspond to cases such as the double burst GRB 110709B \citep{2012ApJ...748..132Z} or the two-shock burst GRB 240529A\citep{2024ApJ...976L..20S}. Thus, the first group of BAT peaks represents the prompt emission of the burst, while the second corresponds to the second burst.
The XRT shows three peaks at $\sim T_{0}+340$\,s, $\sim T_{0}+355$\,s, $\sim T_{0}+380$\,s, then a steep decay closely traces the $\gamma$-ray light curve following the brightest one at $\sim T_{0}+380$\,s. The WHITE band light curve binned with 50\,s begins to rise at $T_{0}+135$\,s followed by the U band light curve binned with 20\,s begins a steeper rise at $T_{0}+278$\,s. Furthermore, the optical data in the g, Rc and Ic band also exhibit similar behavior with a time bin of 60\,s simultaneously. The high temporal-resolution optical data clearly delineate the optical evolution, revealing an abrupt change in optical rise that occurs at $\sim T_{0}+278$\,s. The brightest optical peak occurs at $\sim T_{0}+378$\,s in the U band, then shows a normal decay or even a plateau that does not follow X-ray or $\gamma$-ray data. The optical emission precedes the X-ray and $\gamma$-ray emission, with an initial rise at $T_{0}+135$\,s compared to the delayed rise in the XRT and BAT data at $\sim T_{0}+320$\,s, which implies that they may have different origins.

\subsection{Spectral analysis of the prompt emission} \label{subsec:subsection32}
The joint spectral fitting, including XRT (0.3 - 10 keV), BAT (15 - 150 keV) and optical data, was performed with {\tt XSPEC} (version 12.15.0). We sliced the second prompt emission epoch into 5 time intervals aligned with the optical observation intervals (especially simultaneous optical data in g, Rc and Ic band): $T_{0}+249$\,s to $T_{0}+309$\,s, $T_{0}+319$\,s to $T_{0}+379$\,s, $T_{0}+389$\,s to $T_{0}+449$\,s, $T_{0}+601$\,s to $T_{0}+661$\,s, $T_{0}+670$\,s to $T_{0}+900$\,s. Given the suboptimal SNR of less than 3 in the XRT data from $T_{0}+670$\,s to $T_{0}+770$\,s, we extend the time interval to $T_{0}+900$\,s to ensure enough SNR and the optical flux represents the mean flux during this time interval, obtained by linear interpolation. We jointly fit broadband data with the $\chi^{2}$ statistic and model comparison was quantified with the Bayesian information criterion (BIC) \citep{schwarz1978estimating}. Four models with Galactic absorption ($tbabs$) and host galaxy absorption ($ztbabs$) were used here for spectral fitting in X-ray and $\gamma$-ray bands, including power-law (PL), power-law with blackbody (PL+BB), power-law with Band model (PL+Band) and power-law with synchrotron model (PL+SYN). $tbabs$ and $ztbabs$ refer to the Tuebingen-Boulder ISM absorption model\citep{2000ApJ...542..914W} without or with a fixed redshift parameter, respectively, which calculates the cross section for X-ray absorption by the ISM and allows the user to vary just the equivalent hydrogen column. Because the cutoff energy cannot be tightly constrained within the BAT energy range, we employed a simple power-law (PL) model rather than the CPL model. The Galactic equivalent hydrogen column density $N_{H}$ is fixed as $8.22\times 10^{20}$\,cm$^{-2}$ \citep{2013MNRAS.431..394W}. Given the distinct onset times of the optical and X-ray/$\gamma$-ray emissions, we infer that they arise from different physical origins. Consequently, we attempt to employ a two-component model to disentangle their respective contributions to the spectrum. A GRB spectrum can usually be fit with a smoothly joint broken power law known as the Band function \citep{1993ApJ...413..281B} and the traditional model: a non-thermal component with a quasi-thermal component is considered. Given that synchrotron radiation is a natural candidate for the prompt emission mechanism, we also consider this possibility and discuss it in detail in the following section. For optical data, the Milky Way extinction model with $E(B-V)_{\rm{MW}}= 0.0693$ \citep{1999PASP..111...63F} and $R_{V}=3.08$ is adopted. The extinction model of Small Magellanic Cloud model and $R_{V}=2.93$ is adopted for the host galaxy. When fitting the PL+Band and PL+SYN models, we observed that the low-energy spectral index of the Band function and SYN model are strongly coupled with the equivalent hydrogen column density of the host galaxy $N_{\rm H,host}$ in the soft X-rays and the power-law index in the optical band. Consequently, we fixed these parameters:(1)$N_{\rm H,host}$ = $4.9\times 10^{21}$\,cm$^{-2}$ (2) power-law index $\beta$ = 2  according to the fitting result of late-time ($>T_0+4000$\,s) averaged XRT spectrum mentioned below, assuming no significant spectral evolution of the afterglow component. 

Furthermore, we analyzed the BAT ($15-350$ keV) light curve of GRB 110801A from $T_0+320$\,s to $T_0+400$\,s(binned at 0.01\,s) using the continuous wavelet transform method\footnote{\url{https://github.com/giacomov/mvts}} introduced by \cite{2018ApJ...864..163V}. This analysis resulted in a minimum variability timescale of $t_{\mathrm{mvts}} \sim 1.51$\,s. Adopting this value, we then constrained the Lorentz factor $\Gamma$ in this period and the minimum emission radius $R_c$ following the formalism of \cite{2015ApJ...811...93G}:
\begin{equation}
    \Gamma \gtrsim 110 \left( \frac{L_{\gamma}}{10^{51} \, \mathrm{erg \, s^{-1}}} \frac{1+z}{t_{\mathrm{mvts}}/0.1 \, \mathrm{s}} \right)^{1/5}
\end{equation}
where $L_{\gamma}\sim 1.7\times10^{51}\mathrm{erg}\cdot\mathrm{s^{-1}} $ is the $\gamma$-ray luminosity(1 - $10^{4}$ keV) according to the Konus-Wind fit results and
\begin{equation}
R_{\rm c} \simeq 7.3 \times 10^{13} 
\left( \frac{L_{\gamma}}{10^{51}~\mathrm{erg~s^{-1}}} \right)^{2/5}
\left( \frac{t_{\mathrm{mvts}} / 0.1~\mathrm{s}}{1 + z} \right)^{3/5}~\mathrm{cm}.
\end{equation}
We found $\Gamma \gtrsim 88$ and $ R_c \simeq 2.6 \times 10^{14}~\mathrm{ cm}$ for GRB 110801A in this period, which is consistent with the typical value $\sim 10^{14}$\,cm in \cite{2015ApJ...811...93G} for long-duration GRBs. 

\subsection{Synchrotron Model} 
\label{subsec:subsection33}
Since there is no synchrotron model available in {\tt XSPEC}, we incorporated the synchrotron model from \cite{2020NatAs...4..174B} into our analysis by utilizing a table model. We consider the evolution of the electron distribution function under synchrotron cooling alone obeys a Fokker–Planck-type equation:
\begin{equation}
    \frac{\partial}{\partial t}n_e(\gamma, t) = \frac{\partial}{\partial \gamma}C(\gamma)n_e(\gamma, t)+Q(\gamma )
\end{equation}
where $Q(\gamma)$ are a population of electrons accelerated into a PL energy distribution $\propto \gamma^{-p}$  for Lorentz factors $\gamma$ between a minimum value $\gamma_{min}$ and a maximum value $\gamma_{max}$ with a constant index $p$ and are cooled via emission of synchrotron photons:
\begin{equation}
    C(\gamma) = - \frac{\sigma_{\mathrm{T}} B^2}{6 \pi m_{\mathrm{e}} c} \gamma^2
\end{equation}
where $\sigma_{\mathrm{T}}$ is the Thomson cross-section, $m_e$ the mass of an electron, $c$ the speed of light, and $B$ is the magnetic field strength. 

The total observed emission is characterized by seven parameters: $\Gamma$, $B$, $\gamma_{\min}$, $\gamma_{c}$, $\gamma_{\max}$, $p$ and normalization. Following \cite{2020NatAs...4..174B}, we fix $\gamma_{\min}=10^5$ and $\Gamma=88$. We generate synchrotron spectra by varying $p \in [2, 6]$, $B \in [0.01, 100]$\,G, $\gamma_{c} \in [10^{-2}, 10^{2}]\gamma_{\min}$ (covering slow/fast cooling), and $\gamma_{\max} \in [10^{2}, 10^{3}]\gamma_{\min}$ to ensure high-energy cutoffs above the BAT window. The synchrotron model was implemented in XSPEC in the form of a tabulated additive model.

\subsection{Spectral Results and Model Comparison} \label{subsec:subsection34}
As shown in Figure \ref{fig:2}(a), the X-ray to $\gamma$-ray SEDs exhibit clear excesses in the 2–10 keV range when fitted with a simple power-law (PL) model. These excesses can be adequately described by a thermal component or a non-thermal Band component. Additionally, the PL model yields a poor fit when extrapolated to the optical band. Compared to the single-component PL model, our joint spectral fitting reveals that two-component models employing PL+BB or PL+Band provide significantly better fits during the active phases of the X-ray and $\gamma$-ray emission. However, resolving the additional spectral component is challenging during the quiescent phases ($T_{0}+601$\,s to $T_{0}+661$\,s and $T_{0}+670$\,s to $T_{0}+900$\,s) due to its weak contribution. Specifically, for the interval $T_{0}+601$\,s to $T_{0}+661$\,s, the cutoff energy ($E_{\rm cut}$) in the PL+Band model evolves below the lower threshold of the XRT, resulting in a degeneracy between the low-energy spectral index and the dust extinction $E(B-V)_{\rm{host}}$ in the optical band. Consequently, we fixed $E(B-V)_{\rm{host}}=0.106$ (the average of the three preceding epochs) to effectively constrain the Band model. For the interval $T_{0}+670$\,s to $T_{0}+900$\,s, the photon index of the PL component becomes similar to that of the late-time afterglow, making it difficult to disentangle the second component. Furthermore, we tested a physical synchrotron model (PL+SYN) to fit the multi-band data. Apart from the interval 319–379\,s, the PL+Band model yields a fit only slightly superior to that of the PL+SYN model. Notably, the PL+SYN model provides a substantially better fit than the single PL model during the second active phase, indicating that synchrotron radiation is a viable candidate for observed X-ray and $\gamma$-ray emissions.

Table \ref{table:prompt_fit} details the spectral parameters and BIC statistics for the PL, PL+Band, PL+BB, and PL+SYN models. The PL+Band model generally outperforms the PL+BB model, with the exception of the interval from $T_{0}+249$\,s to $T_{0}+309$\,s. During this specific period, the Blackbody temperature falls within the $\gamma$-ray band, leaving the X-ray spectrum dominated by the power-law component; thus, the PL+BB fit becomes virtually indistinguishable from the single PL model. Overall, the PL+Band model not only yields better BIC statistics but also offers a more physically consistent explanation for the excess emission. In the PL+BB scenario, the excess relative to the power-law is confined to the X-ray band, implying that the optical and $\gamma$-ray emissions share a common origin (the PL component). This contradicts the behavior of the light curve, which suggests that the X-ray and $\gamma$-ray emissions likely originate from the same component. In contrast, in the PL+Band model, the Band component accounts for the excess in both the X-ray and $\gamma$-ray bands, while the optical emission is dominated by the PL component. This interpretation aligns well with the temporal evolution observed in the light curves.

Furthermore, the evolution of $E(B-V)_{\rm{host}}$ derived from the PL+BB fitting shows an unphysical increase over time before decreasing. In contrast, the $E(B-V)_{\rm{host}}$ derived from the PL+Band model remains constant within the error margins before dropping to the late-time afterglow level. While a decrease in $E(B-V)_{\rm{host}}$ is naturally explained by dust destruction due to high-energy UV radiation \citep{10.1093/mnras/stad538}, an increase lacks a reasonable physical interpretation, further favoring the PL+Band model. Incorporating the double-burst hypothesis, we attribute the PL component to the afterglow of the first burst (dominating the optical band) and the Band component to the second burst's prompt emission (dominating the X-ray and $\gamma$-ray bands). 

 Although the statistical performance of the PL+SYN model is slightly inferior to that of the  PL+Band model, it is also worth noting that the PL+SYN model exhibits spectral behavior consistent with that of the PL+Band model. By interpreting the second burst's prompt emission within the synchrotron framework, we conclude that the X-ray and $\gamma$-ray flux likely originate from slow-cooling synchrotron emission during the second active phase.

\begin{figure}[htbp]
\centering
    \subfloat[]{
        \label{sub.1}
        \includegraphics[width=0.48\linewidth]{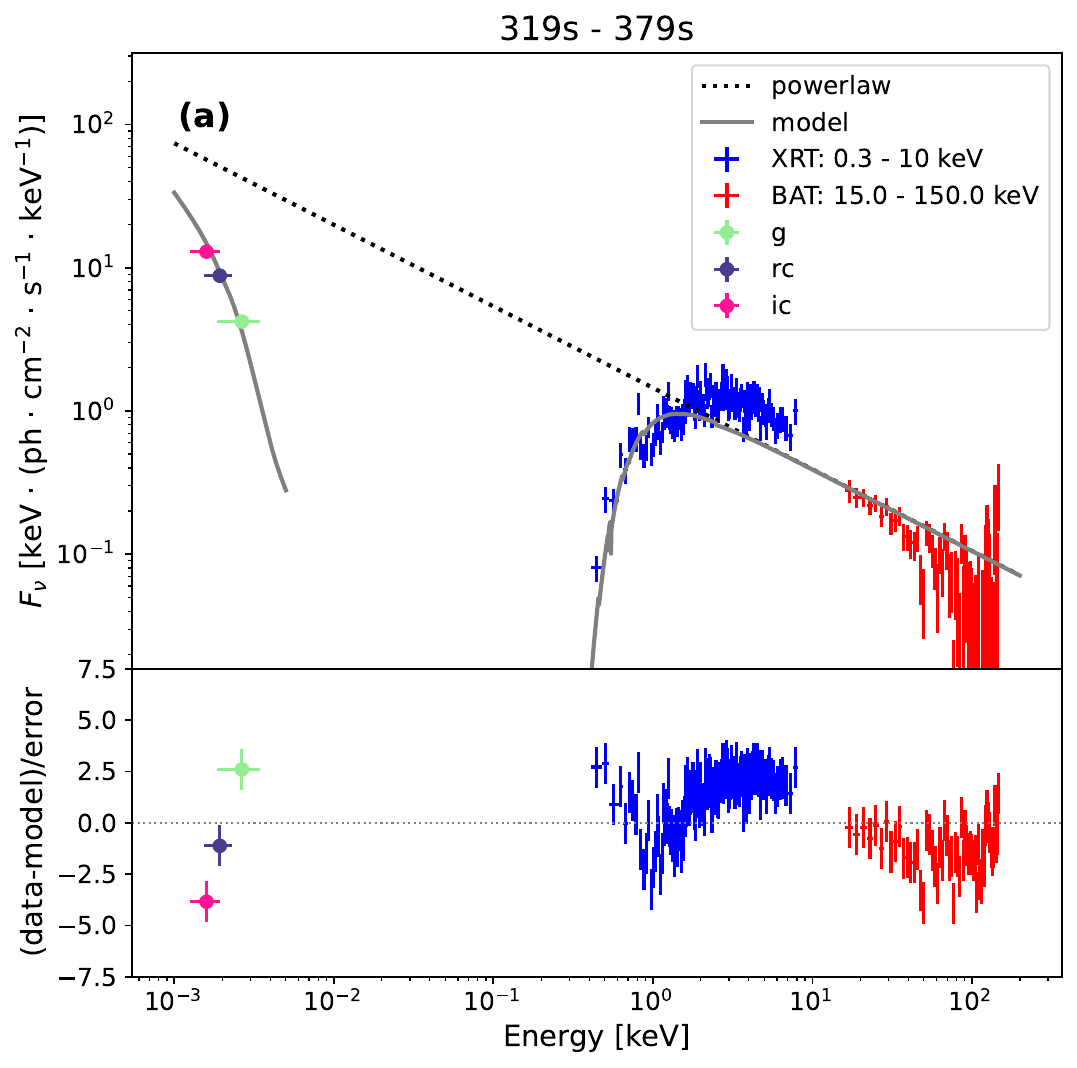}
    }
    \hfill
    \subfloat[]{
        \label{sub.2}
        \includegraphics[width=0.48\linewidth]{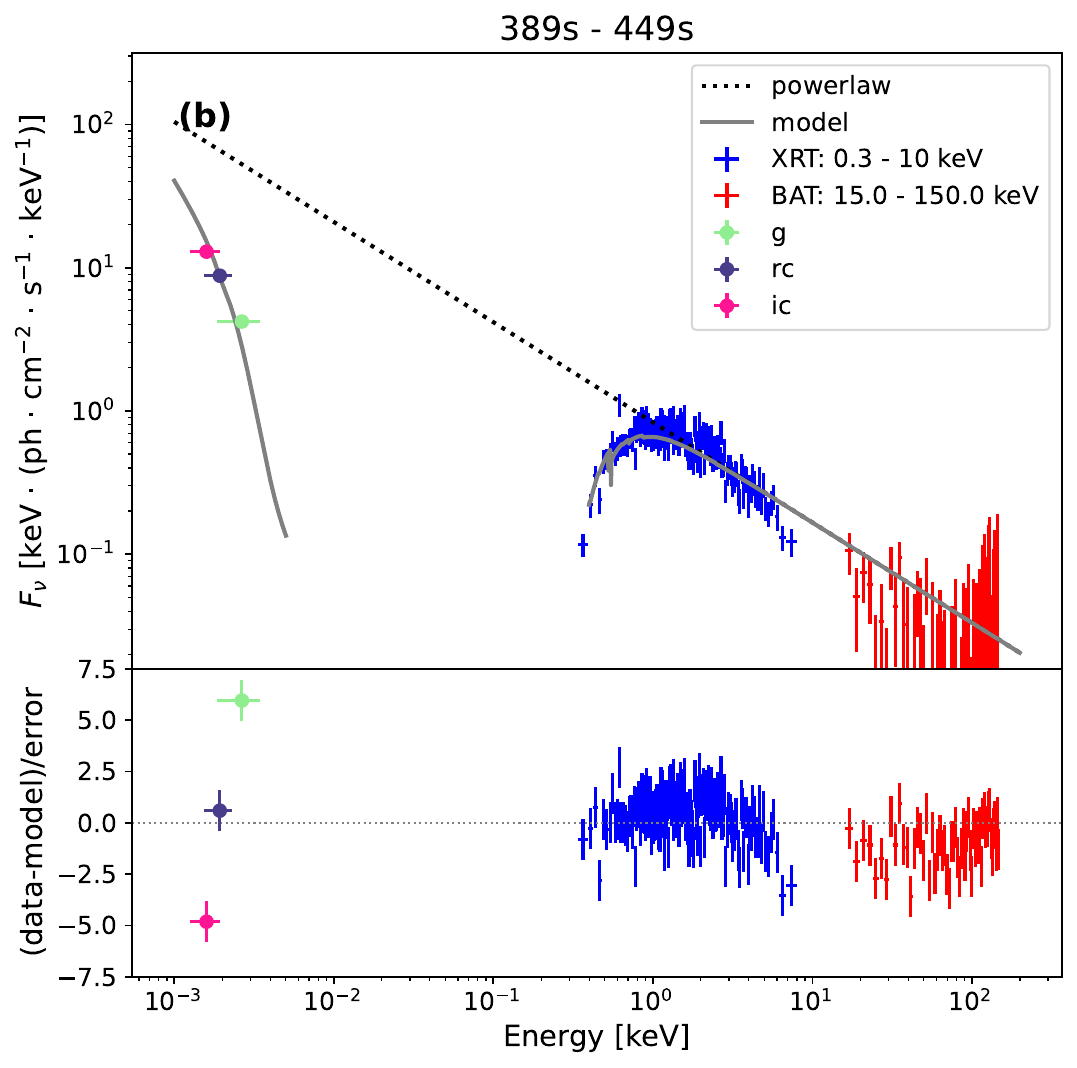}
    }\\
    \subfloat[]{
        \label{sub.3}
        \includegraphics[width=0.48\linewidth]{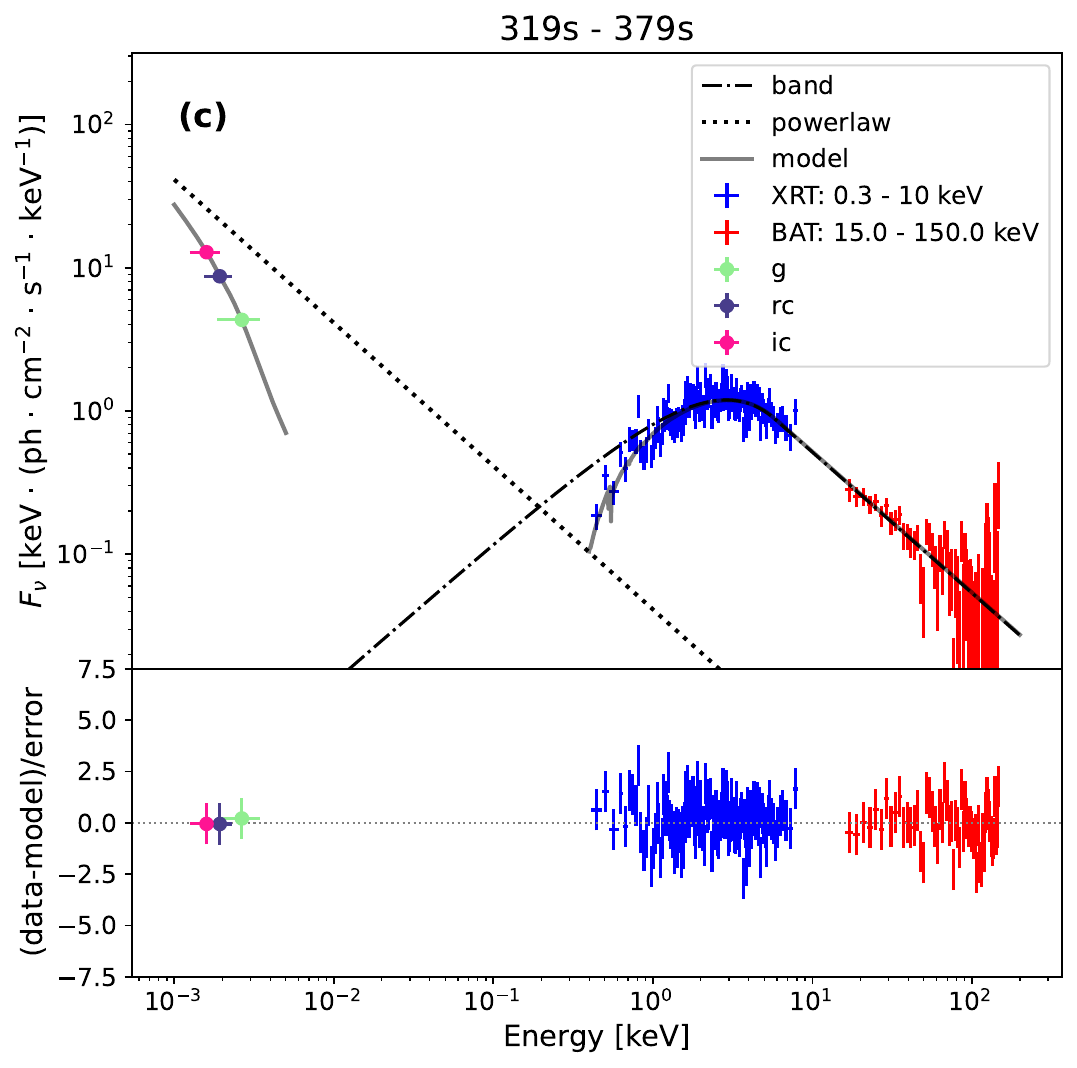}
    }
    \hfill
    \subfloat[]{
        \label{sub.4}
        \includegraphics[width=0.48\linewidth]{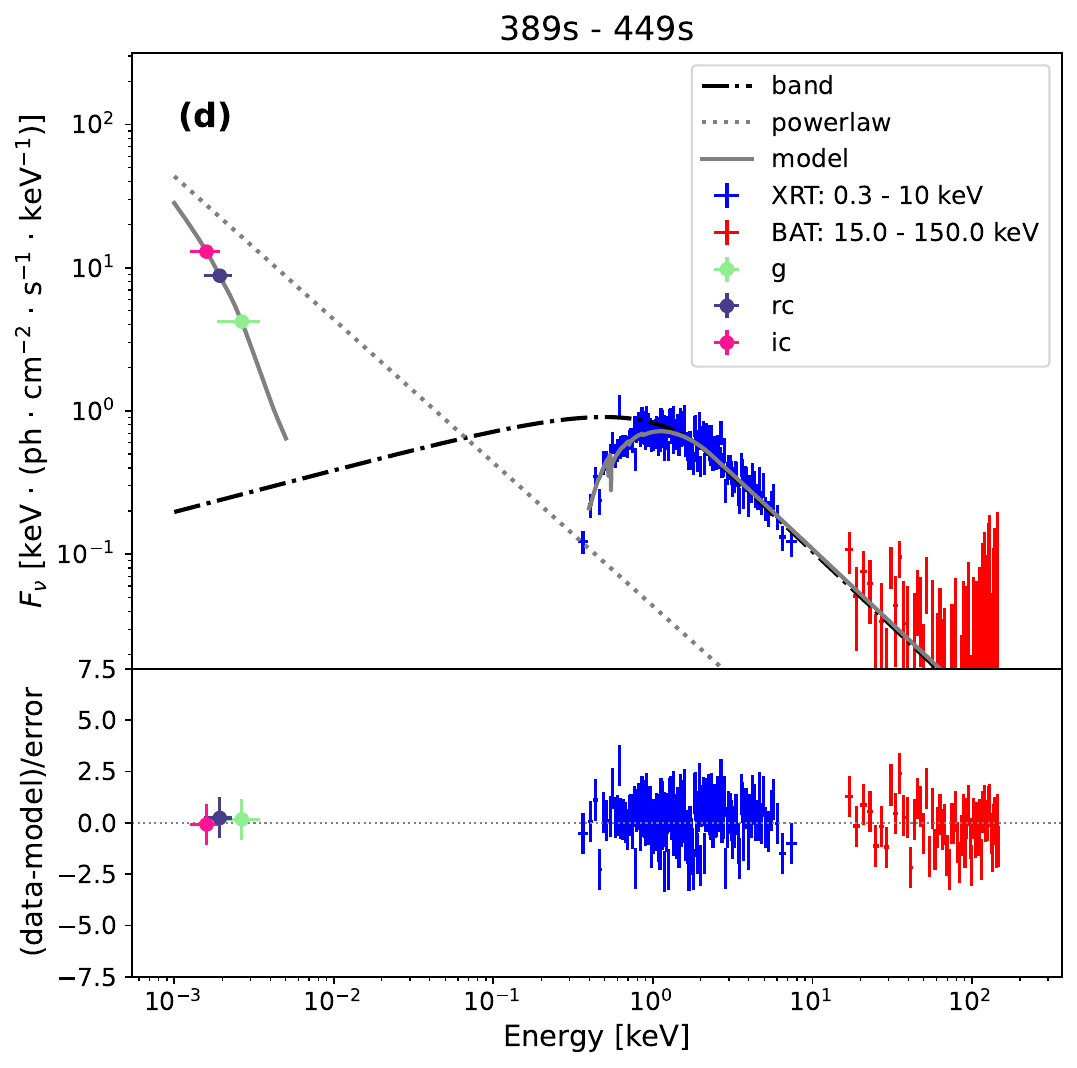}
    }
\caption{The SEDs of GRB 110801A prompt emission. (a)(b) PL only: spectral fitting of the 319--379 s and 389--449 s interval. (c)(d) PL with Band: spectral fitting of the same intervals.}
\end{figure}

\begin{figure}[htbp]
\ContinuedFloat
\centering
    \subfloat[]{
        \label{sub.5}
        \includegraphics[width=0.48\linewidth]{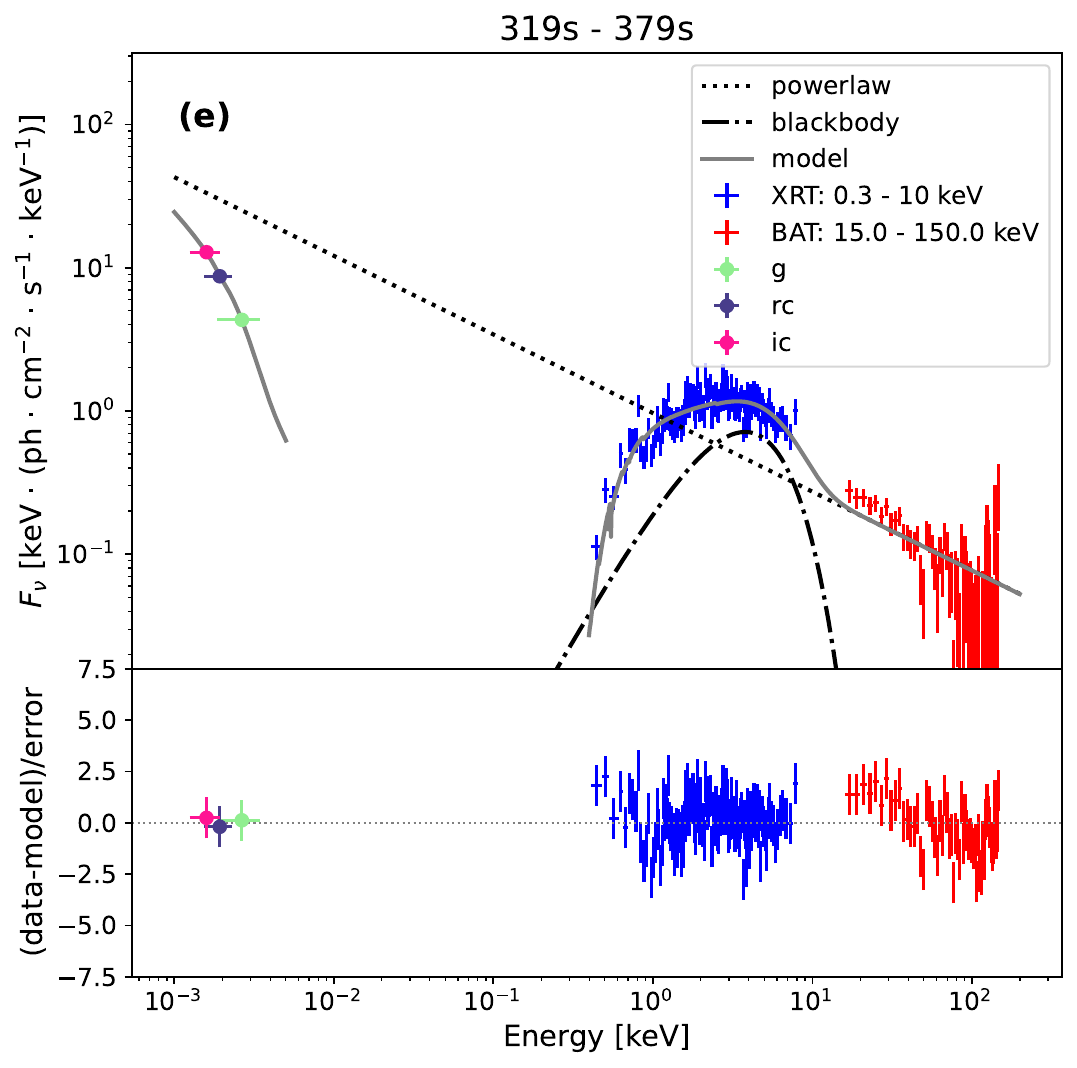}
    }
    \hfill
    \subfloat[]{
        \label{sub.6}
        \includegraphics[width=0.48\linewidth]{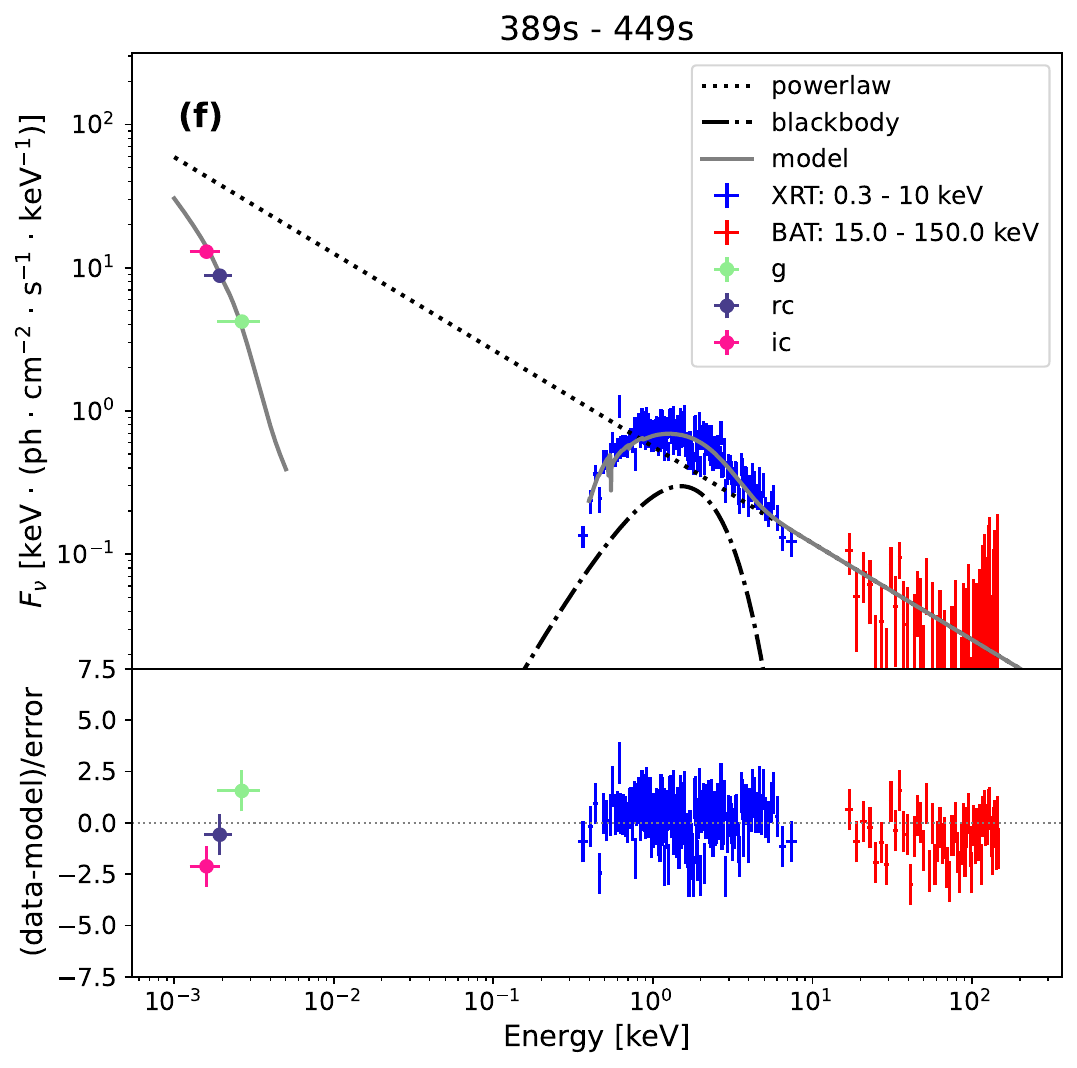}
    }\\
    \subfloat[]{
        \label{sub.7}
        \includegraphics[width=0.48\linewidth]{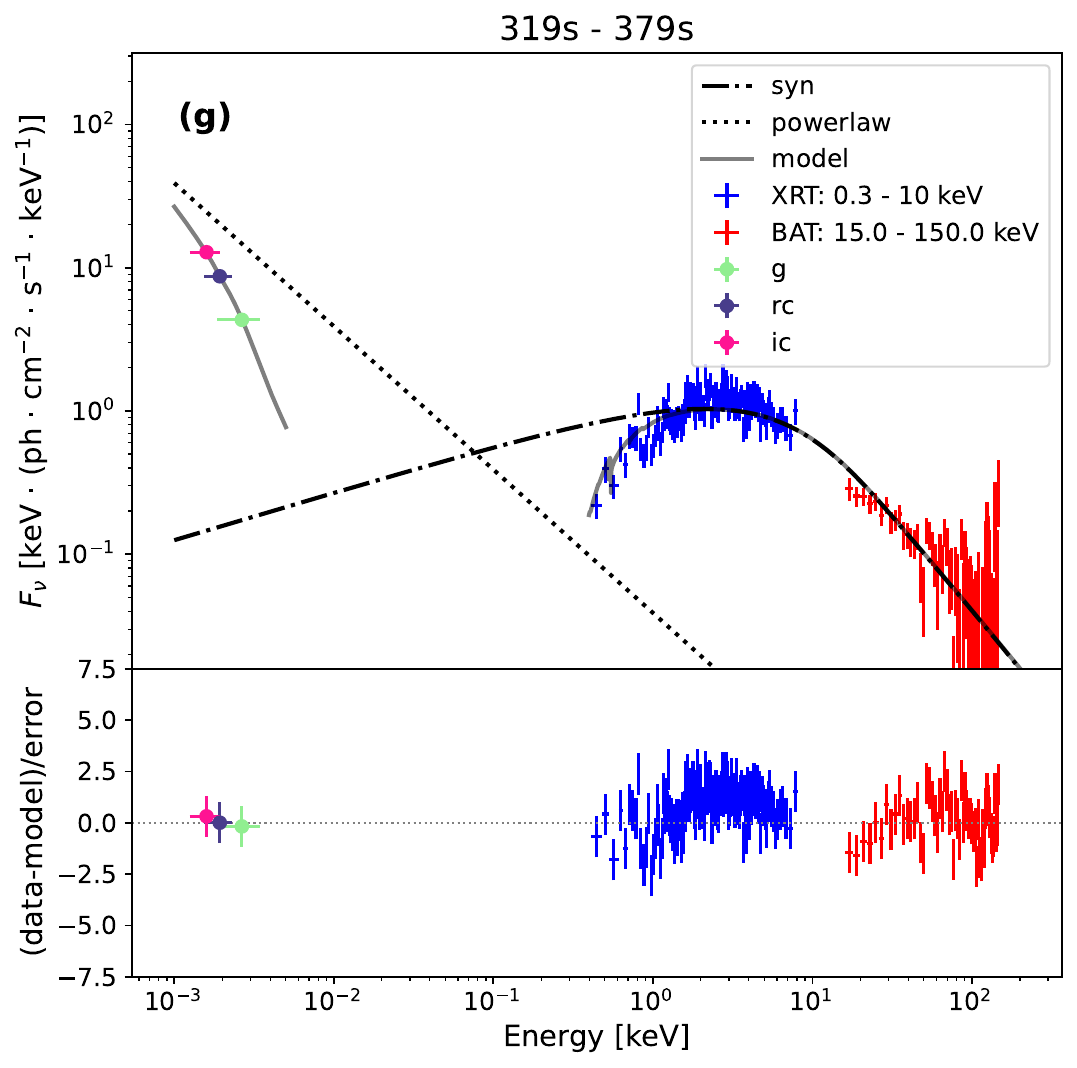}
    }
    \hfill
    \subfloat[]{
        \label{sub.8}
        \includegraphics[width=0.48\linewidth]{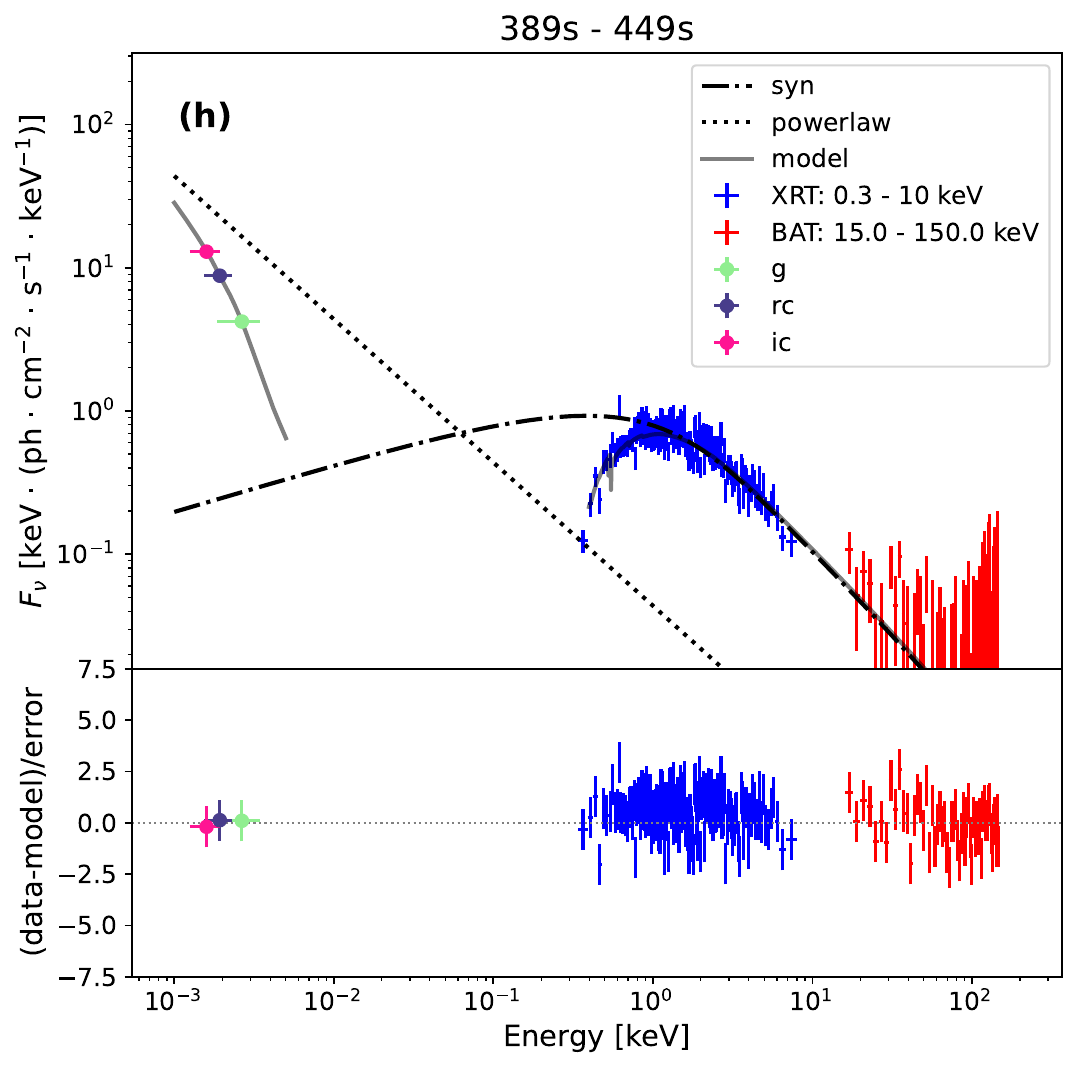}
    }
\caption{(Continued) (e)(f) PL with BB model. (g)(h) PL with SYN model. Dotted lines represent the unabsorbed PL component, dash-dot lines represent the unabsorbed Band, BB, or SYN components, and the solid gray line represents the predictions generated by the model.}
\label{fig:2}
\end{figure}

\begin{table}[htbp]
\centering
\resizebox{\textwidth}{!}{%
\begin{tabular}{l|c|c|c|c|c|c} 
\hline \hline
 Model & Parameters & Episode 1 & Episode 2 & Episode 3 & Episode 4 & Episode 5\\ 
\hline
\multirow{2}{*}{\shortstack[l]{Time range \\ (from $T_0$)}} & Start time (s) & $249$ & $319$ & $389$ & $601$ & $670$ \\
 & End time (s) & $309$ & $379$ & $449$ & $661$ & $900$ \\
\hline
\multirow{7}{*}{Powerlaw+Band} 
 & $\alpha$ & $0.30^{+0.29}_{-0.40}$ & $-0.03_{-0.12}^{+0.12}$ & $-0.71_{-0.15}^{+0.20}$ & $0.26^{+0.34}_{-0.44}$  \\
 & $\beta$ & $-1.66^{+0.09}_{-0.09}$ & $-1.98_{-0.04}^{+0.03}$ & $-2.05_{-0.06}^{+0.06}$ & $-3.77_{-0.41}^{+0.44}$  \\
 & $E_{c}$(keV) & $0.54^{+0.17}_{-0.08}$ & $2.97_{-0.36}^{+0.48}$ & $1.70_{-0.41}^{+0.41}$ & $0.06_{-0.01}^{+0.02}$  \\
 & $K_{Band}$  & $2.36^{+9.52}_{-2.13}$ & $0.99_{-0.44}^{+0.75}$ & $0.06_{-0.03}^{+0.12}$ & $\log_{10}{K_{Band}}=3.3_{-1.7}^{+1.2}$   \\
& $K_{Powerlaw}$  & $1.74^{+0.23}_{-0.19}\times10^{-2}$ & $4.12_{-0.41}^{+0.49}\times10^{-2}$ & $4.35_{-0.37}^{+0.42}\times10^{-2}$ & $3.49_{-0.14}^{+0.10}\times10^{-2}$ \\
& $E(B-V)_{\rm{host}}$ & $0.089^{+0.016}_{-0.014}$ & $0.099_{-0.014}^{+0.015}$ & $0.109_{-0.012}^{+0.012}$ &0.106 \\
 & BIC & 125 & 209 & 251 & 97 \\
\hline
\multirow{7}{*}{Powerlaw+BB}
& $N_{H,host}$($10^{22}$cm$^{-1}$) & $0.73^{+0.29}_{-0.23}$ & $2.55_{-0.34}^{+0.37}$& $0.48_{-0.14}^{+0.29}$ & $0.16^{+2.04}_{-0.13}$  \\
 & $kT$(keV) & $172.04^{+20.81}_{-49.01}$ & $1.34_{0.06}^{+0.07}$ & $0.53_{-0.06}^{+0.06}$ & $3.68_{-3.61}^{+11.47}$  \\
 & $K_{BB}$ & $1.07^{+0.75}_{-0.73}$ & $0.08_{-0.01}^{+0.01}$ & $1.38_{-0.04}^{+0.03}\times10^{-2}$ & $1.86_{-1.53}^{+1.55}\times10^{-3}$  \\
 & $\alpha_{PL}$  & $1.74^{+0.02}_{-0.02}$ & $1.54_{-0.06}^{+0.07}$ & $1.67_{-0.02}^{+0.03}$ & $1.94_{-0.02}^{+0.02}$   \\
& $K_{Powerlaw}$  & $0.11^{+0.01}_{-0.01}$ & $0.96_{-0.04}^{+0.04}$ & $5.60_{-0.40}^{+2.60}\times10^{-3}$ & $0.05_{-0.01}^{+0.01}$ \\
& $E(B-V)_{\rm{host}}$ & $0.115^{+0.014}_{-0.015}$ & $0.145_{-0.012}^{+0.012}$ & $0.174_{-0.019}^{+0.08}$ &$0.100_{-0.014}^{+0.013}$ \\
 & BIC & 123 & 260 & 296 & 110 \\
\hline
\multirow{8}{*}{Powerlaw+SYN}
&$p$&$2.8_{-0.6}^{+2.0}$& $3.5_{-0.2}^{+0.2}$&$3.1_{-0.3}^{+0.2}$&$5.2_{-1.2} ^{+0.6}$ \\
&$\log_{10}{B}$(G)&$-0.24_{-1.41}^{+0.37}$&$-0.52_{-0.04}^{+0.04}$&$-1.30_{-0.06} ^{+0.08}$&$-1.92_{-0.06}^{+0.12}$ \\
&$\log_{10}{\gamma_c}$&$1.25_{-2.60}^{+0.54}$&$-1.60_{-0.28}^{+0.42}$&$-1.27_{-0.49}^{+0.63}$&$0.25_{-0.67}^{+0.63}$ \\
&$\log_{10}{\gamma_{\max}}$&$-2.44_{-0.36}^{+0.30}$&$-2.54_{-0.31}^{+0.34}$&$-2.52_{-0.34}^{+0.34}$&$-2.49_{-0.33}^{+0.34}$ \\
&$\log_{10}{K}$&$-5.23_{-0.52}^{+1.91}$&$-3.51_{-0.26}^{+0.19}$&$-2.84_{-0.36}^{+0.28}$&$-2.65_{-0.47}^{+0.43}$ \\
&$K_{Powerlaw}$& $1.72_{-0.19}^{+0.21}\times10^{-2}$&$3.90_{-0.38}^{+0.40}\times10^{-2}$ &$4.37_{-0.44}^{+0.48}\times10^{-2}$ &$3.41_{-0.36}^{+0.36}\times10^{-2}$ \\
&$E(B-V)_{\rm{host}}$ & $0.094_{-0.015}^{+0.016}$&$0.094_{-0.013}^{+0.013}$&$0.109_{-0.014}^{+0.014}$&$ 0.113_{-0.015}^{+0.016}$ \\
&BIC& 131 &309& 252& 108 \\
\hline
\multirow{5}{*}{Powerlaw}
&$N_{H,host}$($10^{22}$cm$^{-1}$) &$0.77_{-0.24}^{+0.27}$& $3.71_{-0.47}^{+0.35}$ & $0.81_{-0.10}^{+0.11}$&$0.06_{-0.05}^{+0.11}$&$0.137_{-0.10}^{+0.19}$\\
& $\alpha_{PL}$& $1.74_{-0.02}^{+0.02}$&$1.67_{-0.02}^{+0.03}$&$1.70_{-0.01}^{+0.01}$&$1.94_{-0.02}^{+0.02}$&$1.99_{-0.03}^{+0.03}$\\
&$K_{Powerlaw}$&$0.11_{-0.01}^{+0.01}$&$1.45_{-0.04}^{+0.04}$&$0.83_{-0.02}^{+0.02}$&$0.05_{-0.01}^{+0.01}$&$0.03_{-0.01}^{+0.01}$\\
& $E(B-V)_{\rm{host}}$ &  $0.112_{-0.014}^{+0.014}$&$0.216_{-0.006}^{+0.007}$&$0.254_{-0.014}^{+0.015}$&$0.102_{-0.011}^{+0.012}$&$0.089_{-0.023}^{+0.025}$\\
& BIC & 114 &633&418&94&91\\
\hline
\end{tabular}
}
\caption{Spectral Fitting Parameters for GR110801A. $N_{H,host}$ represents the equivalent hydrogen column of host galaxy. The parameter K with different subscripts represents the normalization of corresponding models. $E(B-V)_{\rm{host}}$ represents the dust extinction of host galaxy. $\alpha$, $\beta$ and $E_{c}$ represent the first powerlaw index, the second powerlaw index and the characteristic energy in Band model. $kT$ represent the blackbody temperature in BB model. $\alpha_{PL}$ represent the powerlaw index in PL model. $p$, $\gamma_{c}$, $\gamma_{\max}$ and $B$, represent the powerlaw index of energy distribution, the synchrotron cooling Lorentz factor and the maximum Lorentz factor in electrons acceleration and the magnetic field strength respectively.  BIC represents the Bayesian Information Criterion, where a lower value suggests a statistically preferred model.}
\label{table:prompt_fit}
\end{table}

\section{AFTERGLOW EMISSION ANALYSIS} \label{sec:afterglow}
\subsection{Temporal behavior of the afterglow}
\label{subsec:subsection41}
The afterglow emission is typically characterized by power-law dependencies in both the temporal and spectral regimes with $F_{\nu}=t^{-\alpha}\nu^{-\beta}$. UV/optical light curves of late afterglow after $T_0+4000$\,s were phenomenologically fitted with a simple power-law(SPL). Since the BAT light curve starts to rise at $T_0-50$\,s, reaches the peak at $T_0$ and the $\gamma$-ray emission from $T_0-50$ to $T_0+80$\,s could be the prompt emission of the first burst, the $T_0-50$\,s is treated as the true start time of the first burst in the afterglow modeling(i.e., the launch time of the jet). All times are given from $T_0$ to $T_0+T$ unless otherwise stated. The best fitted SPL model of UV/optical light curves in WHITE, U, B, V and W1 band gives $\alpha_{3} = 1.1\pm0.1$ for late-time afterglow. We also treat the optical data from $T_0+80$ to $T_0+1000$\,s as the afterglow emission of the first burst and fit them with a smoothly broken power-law(SBPL), which has been widely used to fit afterglow light curves for both the rising and decay phases (e.g., \citealp{2007ApJ...670..565L,2012ApJ...758...27L,2015ApJS..219....9W,2018ApJ...859..163H, 2021MNRAS.506.4163L}). The SBPL can be written as :

\begin{equation}
f(t) = A \left[\left(\frac{t}{t_{b}}\right)^{\alpha_{{1}}\Delta}+\left(\frac{t}{t_{b}}\right)^{\alpha_{2}\Delta}\right]^{-1/\Delta}
\end{equation}

where $A$ is the normalization constant, $\alpha_{1}$ and $\alpha_{2}$ are the temporal indices of power-law segments, $t_{b}$ is the break time,  $\Delta$ is the sharpness parameter of the break. During the early afterglow phase, the WHITE-band light curve exhibits a steep rise with $\alpha_{1}=-2.5\pm0.4$, while the U-band shows an even steeper rise with $\alpha_{2}=-6.5^{+1.0}_{-1.3}$ after $T_0+250$\,s, which may indicate a dominated RS component along with a FS. Hence, to account for the behavior of the early afterglow in the brightening phase, both the RS and FS components should be considered.

\subsection{The spectral analysis of the afterglow } \label{subsec:subsection42}
The epoch from $T_0+4500$ to $T_0+58500$\,s is selected to construct the broad band SED since we have the best multi-band coverage data and high SNR. The XRT data from this epoch were collected to create the X-ray spectrum. In section 4.1, the UVOT UV/optical light curves in the decay phase can be well described by a power-law decay with a temporal decay index $\alpha_3=1.1\pm0.1$. Hence, the UV/optical data were interpolated to T0 + 13400 s with the equivalent photon arrival time of $\sim13400$\,s in this epoch to fit the broad-band SEDs. Due to being potentially affected by Lyman-$\alpha$ and the Lyman limit in the UVW1 and U bands, and the uncertainty in defining the effective wavelength of the broad WHITE band, we utilized only the V and B bands for SED fitting. The V and B band fluxes used to construct the SEDs are $1.16\times10^{-4}$\,Jy and $7.92\times10^{-5}$\,Jy, respectively, with adopted uncertainties of 10\%. 

\begin{figure}[!hb] 
  \centering  
  \includegraphics[width=0.75\linewidth]{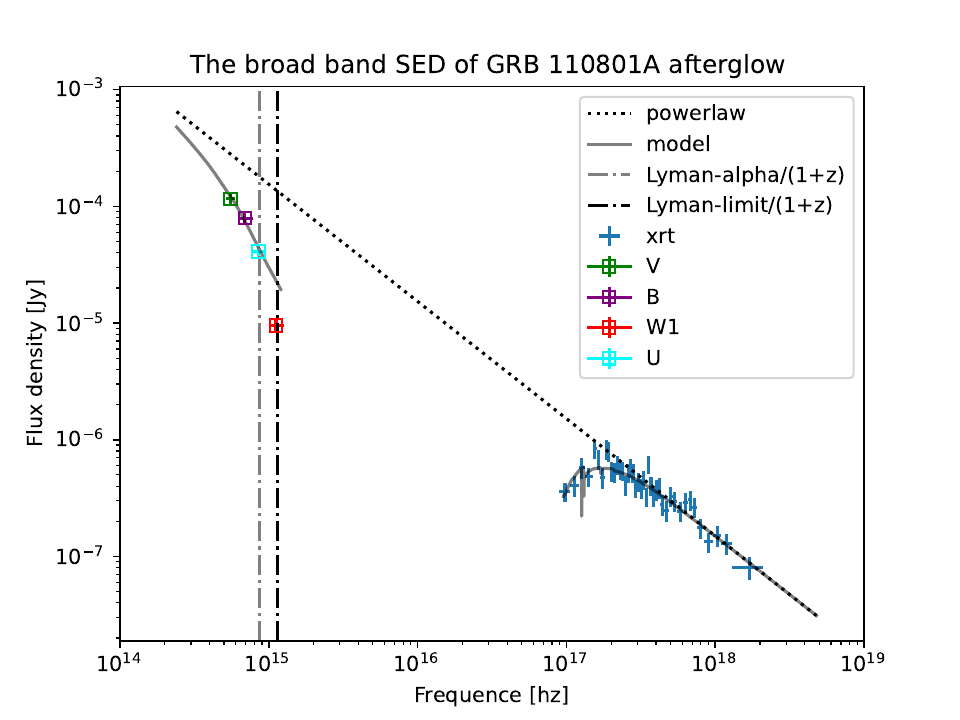}
  \caption{The broad band SED of GRB 110801A afterglow from $T_0+4500$ to $T_0+58500$\,s with the equivalent photon arrival time of $\sim13400$\,s after the BAT trigger. Blue points are observed data of XRT from 0.3 keV to 10.0 keV. Colorful points are the observed data in different UV/optical bands. Gray solid lines represent the predicted absorbed model. Gray and black dash-dot lines represent Lyman-$\alpha$ and Lyman-limit lines. Dashed line is the unabsorbed power-law model with a spectral index of $2.00$.}
  \label{Fig:3}
\end{figure}

The broad band SED was fitted with an absorbed PL model $F_\nu\propto\nu^{-\beta}$. The Milky Way extinction model with $E(B-V)_{\rm{MW}}= 0.0693$ \citep{1999PASP..111...63F} and $R_{V}=3.08$ is adopted. For the host, the extinction model of the Small Magellanic Cloud model is adopted. The best fitted result gives the spectral index $\beta=2.00\pm0.05$, the $E(B-V)_{\rm{host}}$ of the host galaxy is $0.07\pm0.03$ and the equivalent hydrogen column density of the host galaxy $N_{\rm H,host}=4.9^{+2.0}_{-1.9}\times 10^{21}$\,cm$^{-2}$  which is shown in Figure \ref{Fig:3}.

During the {\tt XSPEC} fitting, the $tbabs$ and $ztbabs$ models are set the same as mentioned above. We attempted to fit the data with both SPL and CPL model; however, we cannot use the CPL model here because the cutoff is above the X-ray band and we do not have $\gamma$-ray band detection at this epoch.

\subsection{Afterglow modeling} \label{subsec:subsection43}
In the double-burst scenario, the optical light curve is characterized by a steep rise to a peak around $T_0+378$\,s corresponding to the afterglow of the first burst. The beginning of steep optical rise is not observed by the UVOT, but it can be constrained that the rise starts before the first observation at $T_0+135$\,s in the WHITE band. The steep rise observed in the light curve during the transition from the WHITE band to the U band at $T_0+278$,s can constrain the onset of the RS. The peak reaches U = 16.6 AB mag and following the peak, UV/optical light curves decay monotonically without significant color evolution. After $7\times10^{4}$\,s, a further steepening break in g, Rc and Ic band is observed, probably caused by a jet break.

Meanwhile, the X-ray data show a different temporal behavior. The X-ray light curve begins to rise at $T_0+320$\,s, consistent with the behavior of the $\gamma$-rays. However, this is different from the optical evolution, which shows an onset at $T_0+135$\,s in the WHITE band and a steep rise at $T_0+278$\,s in the U band. After the peak around $T_0+380$\,s, the X-ray light curve shows a monotonic decay and turns into a normal decay with a temporal index of -1.2 at late-time which is similar to optical decay. Subsequently, after $10^{5}$\,s, a comparable break also appears, likely attributed to a jet break seen in the optical bands.

As the relativistic outflow of the GRB propagates into the circumburst medium, it drives a pair of relativistic shocks, i.e., FS and RS components. Initially, the light curve is dominated by the FS emission characterized by an onset in the WHITE band. The RS soon takes over, characterized by a steep rise in the U band. After the peak in the U band, FS dominates the decay phase. The electrons accelerated by relativistic shocks are capable of producing multi-band emissions via synchrotron radiation and inverse Compton scattering. Numerous computational paradigms for modeling afterglows have been developed. To perform numerical calculations of the radiation contributed from the FS component under the various effects, we use a Python-wrapped FORTRAN package\footnote{\url{https://github.com/mikuru1096/ASGARD_GRBAfterglow}}\citep{2024ApJ...962..115R}. For the calculation of the contribution from the RS in the early afterglow phase, it is referenced from \citep{2013ApJ...776..120Y}. Therefore, in this model, the flux density at a certain time and frequency is:
\begin{equation}
    F_{t, \nu}=F(t,\nu, \Gamma_0,\epsilon_{e,f},\epsilon_{B,f},\theta_j,E_{\rm k,iso},p_{f},\epsilon_{e,r},\epsilon_{B,r},p_{r},n_0,f_{e}),
    \label{eq:fs_rs}
\end{equation}
where $\Gamma_0$ is the initial Lorentz factor, $\epsilon_e$ is the fraction of the shock energy converted into the energy of relativistic electrons, $\epsilon_b$ is the fraction of the shock energy converted into the energy of magnetic field, $\theta_j$ is the half-opening angle of the jet in radians, $E_{\rm k,iso}$ is the isotropic kinetic energy of the jet, $p$ is the electron energy distribution index, $n_0$ is the number density of the circumburst medium. And the additional subscripts $f$ and $r$ represent the FS and RS, respectively.

\begin{figure}[htbp] 
  \centering  
  \includegraphics[width=0.75\linewidth]{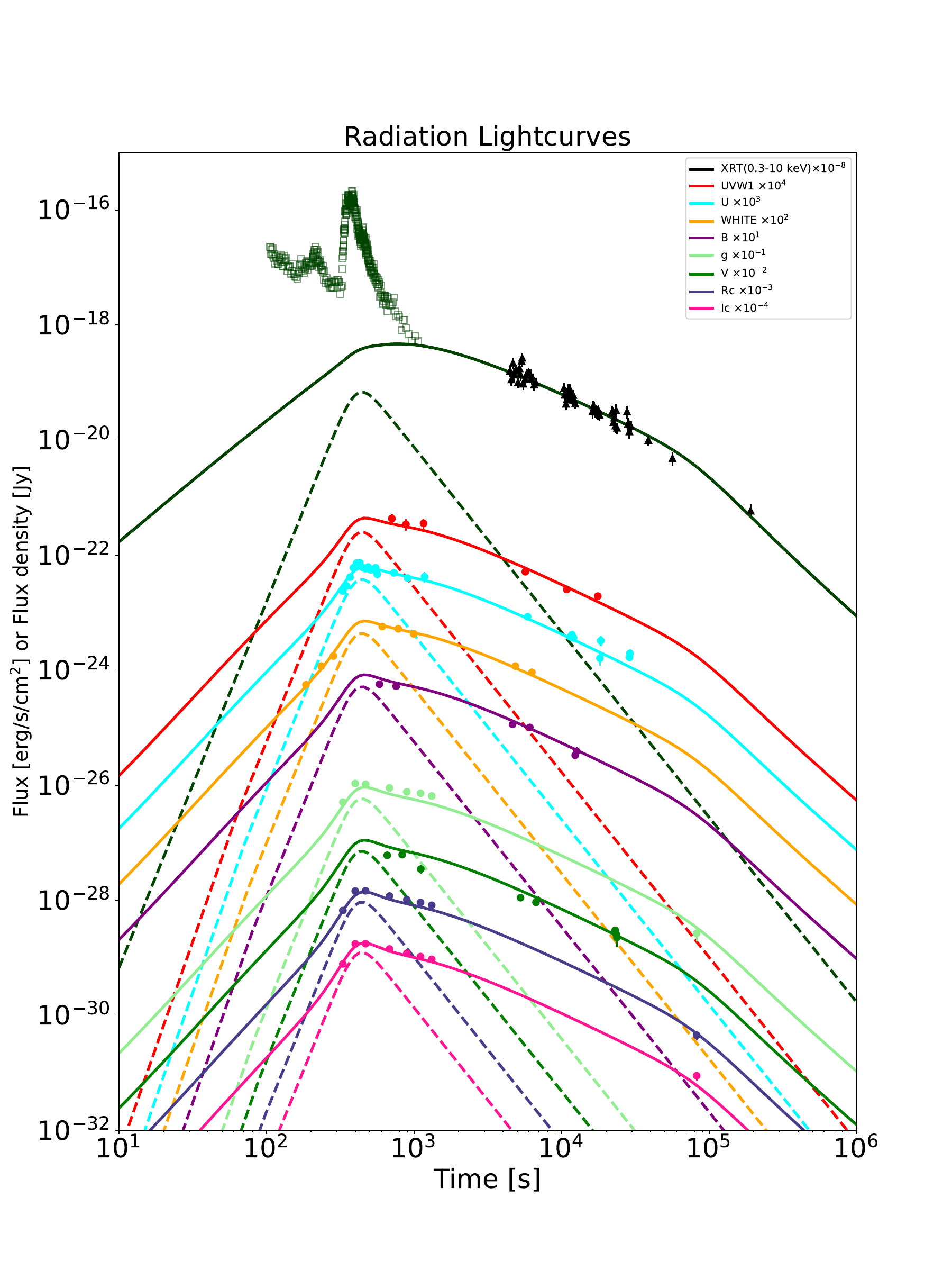}
  \caption{GRB 110801A light curves from XRT, UVOT and OAO data. In this scenario, we considers the optical peak as dominated by RS. The open squares represent XRT data excluded from afterglow fitting. Dashed line represents the RS contribution while solid line represents the sum contribution of all components.}
  \label{Fig:4}
\end{figure}

\begin{figure}[htbp] 
  \centering  
  \includegraphics[width=0.99\linewidth]{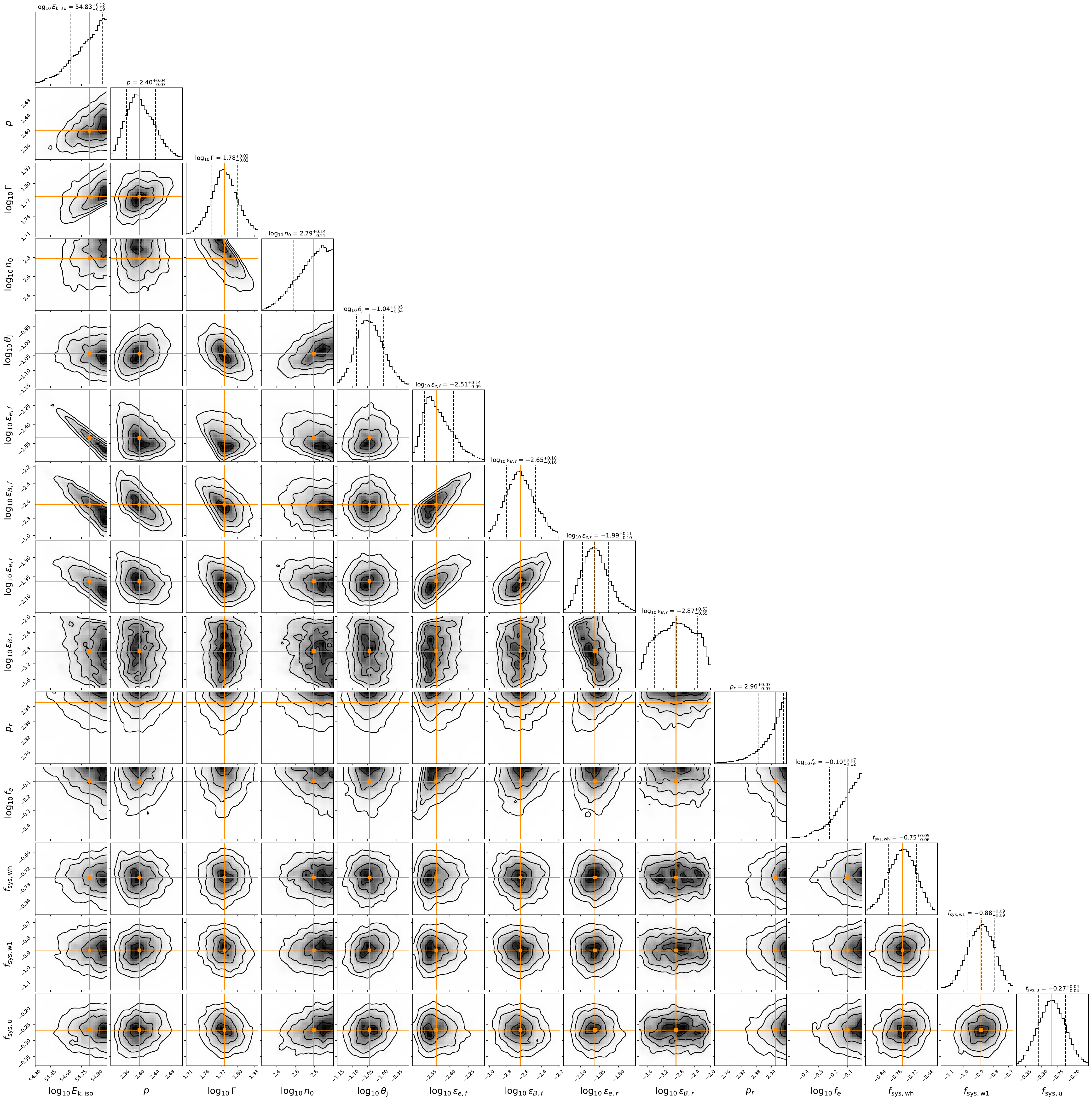}
  \caption{Constrained parameters of the afterglow model for GRB 110801A. Here $\Gamma_0$ is the initial Lorentz factor, $\epsilon_e$ is fraction of energy of relativistic electrons,$\epsilon_b$ is fraction of the energy of magnetic field, $\theta_j$ is the half-opening angle in radians, $E_{\rm k,iso}$ is the isotropic kinetic energy, $p$ is the electron energy distribution index, $n_0$ is the medium number density. And the additional subscripts $f$ and $r$ represent the FS and RS, respectively. $f_{sys,wh}$, $f_{sys,w1}$ and $f_{sys,u}$ are the free parameters to the magnitudes.}
  \label{Fig:5}
\end{figure}

To constrain the physical parameters of our afterglow framework (Eq. \ref{eq:fs_rs}), we employ the {\tt PyMultiNest} package \citep{2014A&A...564A.125B} for Bayesian estimation. The adopted prior distributions are detailed in Table \ref{table:3}. 
We consider a $\gamma$-ray radiation efficiency $\eta_{\gamma}>0.1\%$(the minimum value reported in the statistical study by \cite{2007ApJ...655..989Z}),
 which corresponds to an upper bound of 55 for the prior of $\log_{10}{E_{\rm k,iso}}$. Furthermore, we introduce a log-likelihood term. For a given data point at time $t_i$ in a band with central frequency $\nu_i$, the log-likelihood is calculated:
\begin{equation}
    \ln \mathcal{L}_i=-\frac{1}{2} \frac{(F_{t, \nu} - F_{t, \nu,\rm obs})^2}{\sigma_i^2+f_{sys}^2F_{t, \nu,\rm obs}^2}.
\end{equation}
where $f_{sys}$ is a free parameter to characterize the systematic error. Furthermore, considering the difficulty in accurately calculating the extinction for the WHITE, U, and W1 bands, we introduced a free parameter to the magnitudes of each of these three bands. In the double-burst scenario, the first ejecta initially propagate through the medium while decelerating. The second ejecta, launched approximately during the active phase of the second pulse, moves at a constant velocity because the first ejecta has already swept up most of the ambient material. Consequently, it eventually catches up and collides with the first ejecta. By interpolating the numerical calculation results for the radius evolution of the first ejecta, we derive that the second ejecta, moving at a constant velocity, would catch up with the first ejecta approximately 370 s in the observer frame. We model the energy extraction from the catch-up process as the central engine could operate briefly and inject a stratified ejecta with a distribution of the bulk Lorentz factor, usually defined as the ejecta mass $M(>\gamma) \propto\gamma^{-s}$ moving with a Lorentz factor greater than $\gamma$ \citep{1998A&A...333L..87D,2001ApJ...552L..35Z,2006ApJ...642..354Z}. For this distribution energy injection form, the energy of the shell $E$ with Lorentz factor $\gamma$ is added to the blastwave when the bulk Lorentz factor is decelerated to $\gamma$ and we can derive:
\begin{equation}
   E=E(>\Gamma) \propto\Gamma^{1-s} 
\end{equation}
Furthermore, we can transform this energy injection form into a continuous energy injection equivalently in the form of $L(t)=L_{0}(\frac{t}{t_{s}})^{-q}$, where $q$ is the energy injection
index via the $q-s$ relation in the ISM model from \cite{2006ApJ...642..354Z}:
\begin{equation}
   s=\frac{10-7q}{2+q}, q=\frac{10-2s}{7+s}
\end{equation}
Here we adopt $q=2$ as typical values and the injected energy $E_{inj}=\int_{t_{s}}^{t_s+\Delta t}L(t)dt\sim10^{53}erg$ via Knous-wind fit result and the flux ratio of two pulses observed in the BAT data (we assume radiation efficiency is $10\%$). We define the start time of energy injection, $t_s$, as the catch-up time. The energy injection duration, which corresponds to the thickness of the catching shell, is given by $\Delta t = max(T_{90}, R_s/2\Gamma_{sec}^{2}c)$, where $\Gamma_{sec}$ is the Lorentz factor of
the second catching ejecta and the energy injection into the external shock is assumed to begin at the radius $R_s$ \citep{2018ApJ...852..136L}. Based on our numerical calculation results, we find that $\Delta t=T_{90}\sim50$\,s. The best simultaneous fit results are shown in Figure \ref{Fig:4} and the parameters are summarized in Figure \ref{Fig:5}. By fitting the light curves, we are able to constrain various parameters for both the FS and RS components. According to the FS and RS models, the modeled light curves reproduce both the multi-band optical data and the late-time X-ray emission well, with all derived parameters constrained within typical ranges. This consistency corroborates the hypothesis suggested by our joint spectral fitting that the optical emission is dominated by the afterglow contribution of the first burst. Furthermore, the initial Lorentz factor $\Gamma_0\sim60$, the half-opening angle $\theta_0\sim0.09$ and the isotropic kinetic energy $E_{k,iso}\sim10^{54.8}$erg. The fit result constrains well the fraction of the forward shock energy converted into the energy of magnetic field $\epsilon_{B,f}=2.3\times10^{-3}$ and the reverse shock energy converted into the energy of magnetic field $\epsilon_{B,r}=1.3\times10^{-3}$.

\section{SUMMARY AND DISCUSSION} \label{sec:summary}
We presented a detailed analysis of the prompt emission and the early afterglow of GRB 110801A, and suggest GRB 110801A is a double-burst GRB triggered by its first burst in $\gamma$-ray emission. Initially, a pulse in the $\gamma$-ray band triggered the BAT. Approximately 130 s after the trigger, the optical light curve began to rise. This onset preceded the second pulse observed in both X-ray and $\gamma$-ray bands, where the flux in the $\gamma$-ray band of the second pulse was comparable to that of the first. These features suggest the presence of two distinct burst episodes. The optical band exhibited a significant change in its temporal behavior. If we assume $T_0-50$\,s as the true onset of the burst, the temporal index of the optical rise steepened from -2.5 to -6.5 between $T_0 + 230$\,s and $T_0 + 280$\,s. Such a transition and rapid rise are attributed to the contribution of the reverse shock of the afterglow. After 1000\,s, both the UV/optical and X-ray light curves show normal decay behaviors. We summarize our main results as follows:

(i) According to simultaneous multi-band observation, we find that GRB 110801A exhibits two comparable bursts in the $\gamma$-ray band before and after the trigger, implying that it may be a double-burst event. The temporal discrepancy between the early optical brightening and the delayed rising in the X-ray/$\gamma$-ray band points to different physical origins. This is consistent with previous literature, where chromatic behaviors between optical and X-ray bands are often observed \citep{2025MNRAS.543.2404R}. The comprehensive data of GRB 110801A make it possible to study the mechanisms of emission and to constrain several parameters of the standard fireball model.

(ii) After the first burst, the {\it Swift}/XRT, {\it Swift}/BAT and OAO data were jointly fitted. We find that an additional Band component is required to account for the excess in the X-ray and $\gamma$-ray bands, and the optical emission can be attributed to a power-law component. Moreover, by constraining the power-law parameters based on the late-time afterglow data, we obtained a reasonable fit. This result indicates that the optical radiation likely originates from the afterglow of the first burst, whereas the X-ray and $\gamma$-ray emissions are more likely associated with the prompt emission of the second burst in the framework of synchrotron mechanism. 

(iii) Considering the transition from a normal rise($\sim t^{2.5}$) to a steep rise($\sim t^{6.5}$) observed in the WHITE and U bands, we adopt the scenario that both the RS and the FS contribute considerably to the afterglow at the early stage. Due to  well-sampled data of the afterglow in the early phase, some physical parameters of RS and FS can be well constrained. The best fitted model shows that the UV/optical light curves are initially dominated by the RS and FS takes over after $\sim T_0+378$\,s. From RS and FS modeling, we have the initial Lorentz factor $\Gamma_0$ $\sim$ 60, the half-opening angle $\theta_j \sim0.09$, isotropic kinetic energy $E_{\rm k,iso}\sim 10^{54.8}$ erg, and $\mathcal{R}_{B}=(\epsilon_{B,r}/\epsilon_{B,f})^{\frac{1}2{}}\approx 0.8$. The posterior distributions of shock parameters like $E_{\rm k,iso}$ and $n_0$ tend to peak at the upper boundaries of their priors due to the degeneracy between $E_{\rm k,iso}$ and $n_0$ in afterglow modeling. Furthermore, the absence of radio-band observations for this burst from effectively constraint $n_0$. While these values are higher than those of typical long GRBs, we note that similar fitting behaviors—where $E_{\rm k,iso}$ and $n_0$ tend toward large values—have been reported in another two-shock event, GRB 240529A \citep{2024ApJ...976L..20S}. Regarding the reverse shock electron index $p_r$, the fitting result indeed hits the prior upper edge. This occurs because we excluded the X-ray data during the reverse-shock-dominated epoch, as we interpret this emission as an X-ray flare in the prompt emission phase. As a result, a few remaining optical data points are insufficient to tightly constrain $p_r$.

In conclusion, with the broadband coverage and high temporal resolution of double-burst GRB 110801A, our analysis reveals that the pulse in the X-ray and $\gamma$-ray bands at approximately 320\,s originates from internal shocks of the sceond burst, while the rapid rise in the optical band originates from the external shock of the first burst and their parameters are constrained. 

\begin{acknowledgments}
We acknowledge the use of the public data from the Swift archive. This work is supported by the National Key R\&D Program of China (grant Nos. 2024YFA1611704 and 2024YFA1611700), the Natural Science Foundation of China (grant Nos. 12225305, 12473049, 12321003, and 12233011), the Strategic Priority Research Program of the Chinese Academy of Sciences (grant No. XDB0550400), the Jiangsu Funding Program for Excellent Postdoctoral Talent (grant No.~2024ZB110), the Basic Research Program of Jiangsu (grant No.~BK20251707), the Postdoctoral Innovation Talents Support Program (No.~BX20250159), the Postdoctoral Fellowship Program (grant No.~GZC20241916), and the General Fund (grant No.~2024M763531) of the China Postdoctoral Science Foundation.

\end{acknowledgments}

\facilities{Swift(BAT, XRT and UVOT)}

\clearpage 
\appendix
In this appendix, we provide detailed information on the observation data and fitting results of the parameters in the prompt and afterglow phases.

\begin{ThreePartTable}
\begin{TableNotes}\footnotesize
    \item[]{\bf Note.} The photometry results presented in this table have not been corrected for Galactic extinction.
\end{TableNotes}
\begin{longtable}[htbp]{c c c c c }
\caption{{\it Swift} UVOT photometric observation\label{table:2}}\\

\hline
$T-T_{0}(\rm s)$ & Exposure(s) & Magnitude & Error & Filter\\ 
\hline
\endfirsthead

\hline
$T-T_{0}(\rm s)$ & Exposure(s) & Magnitude & Error & Filter\\ 
\hline
\endhead

\hline
\endfoot

\hline
\multicolumn{5}{c}{End of the Table}\\
\hline
\insertTableNotes\\
\endlastfoot

135 & 50 & 19.62 & 0.15 & WHITE \\
185 & 50 & 18.81 & 0.08 & WHITE \\
234 & 50 & 18.37 & 0.07 & WHITE \\
278 & 20 & 17.78 & 0.15 & U \\
298 & 20 & 17.56 & 0.13 & U \\
318 & 20 & 17.18 & 0.11 & U \\
338 & 20 & 16.77 & 0.09 & U \\
358 & 20 & 16.57 & 0.08 & U \\
378 & 20 & 16.55 & 0.08 & U \\
398 & 20 & 16.76 & 0.09 & U \\
418 & 20 & 16.81 & 0.09 & U \\
438 & 20 & 16.74 & 0.09 & U \\
458 & 20 & 16.86 & 0.10 & U \\
478 & 20 & 16.85 & 0.10 & U \\
498 & 20 & 16.77 & 0.09 & U \\
512 & 8 & 17.05 & 0.16 & U \\
532 & 20 & 16.09 & 0.09 & B \\
558 & 20 & 17.09 & 0.06 & WHITE \\
607 & 20 & 15.64 & 0.12 & V \\
657 & 20 & 18.28 & 0.22 & UVW1 \\
681 & 20 & 16.99 & 0.11 & U \\
706 & 20 & 16.18 & 0.09 & B \\
731 & 20 & 17.19 & 0.06 & WHITE \\
780 & 20 & 15.61 & 0.12 & V \\
830 & 20 & 18.53 & 0.25 & UVW1 \\
854 & 20 & 17.22 & 0.13 & U \\
945 & 150 & 17.41 & 0.03 & WHITE \\
1059 & 20 & 16.24 & 0.18 & V \\
1109 & 18 & 18.49 & 0.24 & UVW1 \\
1127 & 6 & 17.16 & 0.22 & U \\
4597 & 197 & 17.43 & 0.05 & B \\
4802 & 197 & 18.33 & 0.03 & WHITE \\
5212 & 197 & 17.18 & 0.08 & V \\
5622 & 197 & 19.93 & 0.15 & UVW1 \\
5827 & 197 & 18.36 & 0.07 & U \\
6032 & 197 & 17.56 & 0.07 & B \\
6237 & 197 & 18.60 & 0.04 & WHITE \\
6648 & 197 & 17.38 & 0.14 & V \\
10782 & 886 & 20.70 & 0.11 & UVW1 \\
11389 & 295 & 19.23 & 0.08 & U \\
11692 & 295 & 19.14 & 0.09 & U \\
11996 & 295 & 19.28 & 0.12 & U \\
12301 & 295 & 18.77 & 0.17 & B \\
12565 & 218 & 18.60 & 0.19 & B \\
17515 & 886 & 20.99 & 0.15 & UVW1 \\
18121 & 295 & 20.16 & 0.28 & U \\
18361 & 171 & 19.40 & 0.21 & U \\
23009 & 295 & 18.60 & 0.17 & V \\
23313 & 295 & 18.76 & 0.23 & V \\
23616 & 295 & 18.91 & 0.35 & V \\
28747 & 295 & 20.12 & 0.15 & U \\
29002 & 200 & 19.95 & 0.18 & U \\

\end{longtable}
\end{ThreePartTable}

\begin{table}[htbp]
\centering
\caption{Fitting results of model parameters}
\label{table:3}
\begin{tabular}{ccc}
\hline \hline
Parameter & Prior range & Posterior value \\
\hline
$\log_{10}\Gamma_{0}$           & $[1,2.5]$   & $1.78^{+0.02}_{-0.02}$ \\
$\log_{10}\epsilon_{e,f}$       & $[-5-,-0.1]$ & $-2.51^{+0.14}_{-0.09}$ \\
$\log_{10}\epsilon_{B,f}$       & $[-8,-0.1]$ & $-2.65^{+0.18}_{-0.16}$ \\
$\log_{10}\theta_{j}$ (rad)     & $[-2,0]$    & $-1.04^{+0.05}_{-0.04}$ \\
$\log_{10}E_{\rm k,iso}$ (erg)  & $[51,55]$   & $54.83^{+0.12}_{-0.19}$ \\
$p_{f}$                         & $[2,3]$     & $2.40^{+0.04}_{-0.03}$ \\
\hline
$\log_{10}\epsilon_{e,r}$       & $[-5,-0.1]$ & $-1.99^{+0.11}_{-0.10}$ \\
$\log_{10}\epsilon_{B,r}$       & $[-8,-0.1]$ & $-2.88^{+0.54}_{-0.54}$ \\
$p_{r}$                         & $[2,3]$     & $2.96^{+0.03}_{-0.07}$ \\
\hline
$\log_{10}n_{0}$                & $[-3,3]$    & $2.79^{+0.14}_{-0.21}$ \\
$\log_{10}f_{e}$                & $[-3,0]$    & $-0.10^{+0.07}_{-0.13}$ \\
$f_{sys,wh}$              & $[-3,3]$    & $-0.75^{+0.05}_{-0.05}$ \\
$f_{sys,w1}$              & $[-3,3]$    & $-0.89^{+0.09}_{-0.10}$ \\
$f_{sys,u}$              & $[-3,3]$    & $-0.27^{+0.04}_{-0.04}$ \\
\hline
\multicolumn{3}{l}{\footnotesize $^a$ Uniform prior distribution.} \\
\multicolumn{3}{l}{\footnotesize $^b$ The confidence interval is set to $1\sigma$.}
\end{tabular}
\end{table}

%% For this sample we use BibTeX plus aasjournalv7.bst to generate the

\bibliography{main.bib}{}
\bibliographystyle{aasjournalv7}



\end{document}